\documentstyle[12pt,worldsci,epsfig]{article}

\begin{document}
\vspace*{-1.3cm}
\begin{flushright}OCIP/C-93-08\end{flushright}
\vspace{-0.7cm}
\title{{\bf SINGLE $W$-BOSON PRODUCTION IN $e\gamma$
COLLISIONS}\thanks{This work was supported by the Natural
Sciences and Engineering Research Council of Canada.}
\thanks{Talk presented by S. Godfrey at The 2nd
International Workshop on Physics and Experiment
with Linear $e^+e^-$\ Colliders, Waikoloa,
Hawaii, Apr. 26-30, 1993.}
}
\author{STEPHEN GODFREY and K. ANDREW PETERSON \\
{\em Ottawa Carleton Institute for Physics, \\
 Department of Physics, Carleton University, \\
Ottawa, Ontario CANADA K1S 5B6}}

\maketitle
\setlength{\baselineskip}{2.6ex}

\begin{center}
\parbox{13.0cm}
{\begin{center} ABSTRACT \end{center}
{\small \hspace*{0.3cm}

We studied single $W$ boson production in high energy $e\gamma$
collisions and the sensitivity of various observables to the $WW\gamma$
vertex.  We included gauge boson decay to
final state fermions and all contributions to the same final state.
The contributions of the non-resonant diagrams and
their interference with the resonant gauge boson production diagrams
give significant contributions which should not be neglected.
We present results for $W$ production
at a 500 GeV $e^+e^-$ collider with the  photon
spectra obtained from  a  backscattered laser.  }}
\end{center}

\section{Introduction}

Now that the standard model has been confirmed to the level of
radiative corrections the particle physics community has become
obsessed with finding out what lies beyond.
An approach to quantifying the effects of new physics
is to represent new physics by additional terms in an
effective Lagrangian expansion and then to constrain the
coefficients of the effective Lagrangian by precision
experimental measurements \cite{tgv}.
The bounds obtained on the coefficients
can then be related to possible theories of new physics.

In this spirit a commonly used parametrization of
the trilinear gauge boson  CP invariant
effective Lagrangian is given by\cite{vertex}:
\begin{equation}
{\cal L}_{WW\gamma} =  - ie \left\{ {(W^\dagger_{\mu\nu}W^\mu A^\nu -
W^\dagger_\mu A_\nu W^{\mu\nu} )
+ \kappa_\gamma W^\dagger_\mu W_\nu F^{\mu\nu}
+ {{\lambda_\gamma}\over{M_W^2}} W^\dagger_{\lambda\mu}W^\mu_\nu F^{\nu\lambda}
}\right\}
\end{equation}
where $A^\mu$ and $W^\mu$ are the photon and $W^-$ fields,
$W_{\mu\nu}=\partial_\mu W_\nu - \partial_\nu W_\mu$
and $F_{\mu\nu}=\partial_\mu A_\nu - \partial_\nu A_\mu$.
In the standard model, at tree
level, $\delta\kappa_\gamma \simeq 0 \simeq \lambda_\gamma$ while
radiative corrections from either the standard model or new physics
are typically $\delta\kappa \sim O(10^{-2})$ and
$\lambda \sim O(10^{-3})$.

Although bounds can be extracted from high precision low energy
measurements and measurements at the $Z^0$ pole, there are ambiguities
in the results \cite{tgv}.  In contrast, gauge boson
production at colliders can measure the gauge boson couplings directly
and unambiguously.
At the NLC there are a large number of processes which can measure the
TGV's.  In this paper we restrict ourselves to single $W$ production
at $e\gamma$ colliders using
photon spectra produced from backscattered
lasers\cite{egamma} \cite{otheregamma}.

\section{Calculation and Results}

The Feynman diagrams contributing to the process $e^-\gamma \to \nu
f\bar{f}$ are given in Fig. 1.  The
$WW\gamma$ vertex we are studying contributes via diagram 1a.  To
preserve electromagnetic gauge invariance and to properly take into
account the background processes our calculation includes all the
diagrams of Fig 1.  To obtain the cross sections and distributions we
used the CALKUL helicity amplitude technique \cite{calkul}.
Monte Carlo integration techniques are then used to
perform the phase space integrals.  We treat the photon distributions
as structure functions, $f_{\gamma/e}(x)$, and integrate them with the
$e\gamma$ cross  sections to obtain our results.

\begin{figure}
\begin{center}
\mbox{\epsfig{file=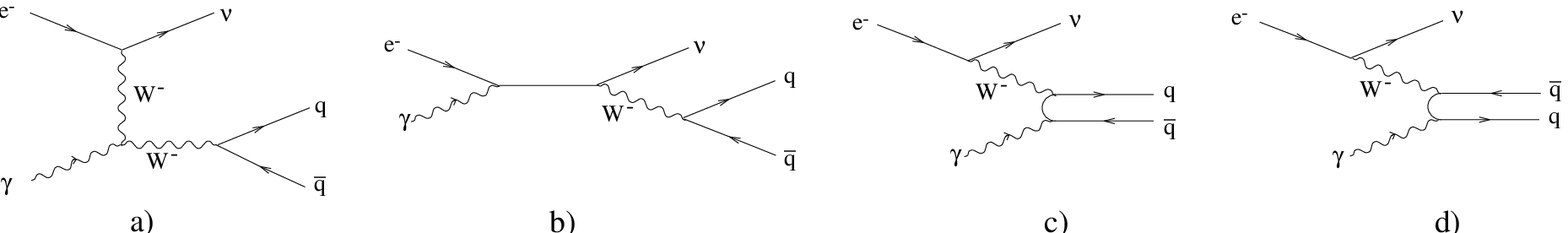,height=2cm}}
\end{center}
\noindent
{\small Fig. 1.  The Feynman diagrams contributing to the process $e\gamma \to
\nu q\bar{q}$. For the process $e\gamma \to \nu_e \mu \bar{\nu}_\mu$
diagram (c) does not contribute. }
\end{figure}

The signal we are studying consists of either, (i) for leptonic $W$ decay,
a high transverse momentum lepton ($p_T$) and
large missing transverse momentum ($\not{p}_T$) due to the neutrinos
from the initial electron beam and from the $W$ decay, or (ii) for hadronic
$W$
decay, two hadronic jets and large missing transverse momentum ($\not{p}_T$)
due to
the neutrino from the initial electron
beam.  In both cases, we require that visible particles in the final state
be at least $10^o$ from the beam direction.  We also imposed the cut;
$\not{p}_T > 10$~GeV.
The signals we consider are therefore
\begin{eqnarray}
& e^- + \gamma \to \mu^- + \not{p} \\
& e^- + \gamma \to j  + j  + \not{p}
\end{eqnarray}
For the hadronic $W$ decay modes we
reconstructed the $W$ boson 4-momentum from the hadronic jets'
4-momentum, imposing the kinematic cut of 75 GeV
$<M_{q\bar{q}}=\sqrt{(p_q +p_{\bar{q}})^2}<$ 85 GeV.
Including the nonresonant diagrams of fig (1c) and (1d) and reconstructing
the $W$ boson in this manner gives different results than from simply
studying the cross sections to real $W$ bosons.

The NLC is envisaged as a very high luminosity collider so that  the
integrated luminosity for a Snowmass year ($10^7$ sec) is expected
to be $\sim 60$ fb$^{-1}$ for a $\sqrt{s}=500$ GeV collider.
Typical cross sections for
the process $e\gamma\to \mu \bar{\nu}_\mu \nu_e$ and
$e\gamma\to W \nu \to q \bar{q} \nu$ at $\sqrt{s}=500$ GeV
are 3.2 pb and 16.6 pb respectively for the backscattered laser mode,
leading to $\sim 10^6$ events per year.
Thus, except for certain regions of phase space,
the errors are not limited by
statistics, but rather by systematic errors.
For cross sections we
assume a systematic error of 5\%.
We combine  the statistical errors in quadrature with the systematic
errors; $\delta^2 = \delta^2_{stat} + \delta^2_{sys}$.

The total cross sections and the angular distributions of the outgoing muon
and reconstructed $W$ are sensitive to anomalous couplings.
At higher energies we can
obtain additional information, especially for $\lambda_\gamma$,
from the $p_T$ spectrum of the outgoing
lepton or the reconstructed $W$.  Angular and
$p_T$ distributions are shown in Fig. 2 for the reconstructed $W$.
The invariant mass distribution
of the
$q\bar{q}$ pair above the $W$ mass also provides useful information as
can be seen in Fig. 3.
For example, integrating the $M_{q\bar{q}}$ spectrum from 300 GeV up,
gives $\sigma= 0.006$~pb
which yields $\sim 400$ events/year.  More importantly, this high
$M_{q\bar{q}}$
region shows a higher sensitivity to anomalous couplings than the $W$ pole
region.

\begin{figure}
\begin{minipage}{.495\linewidth}
\begin{center}
\mbox{\epsfig{figure=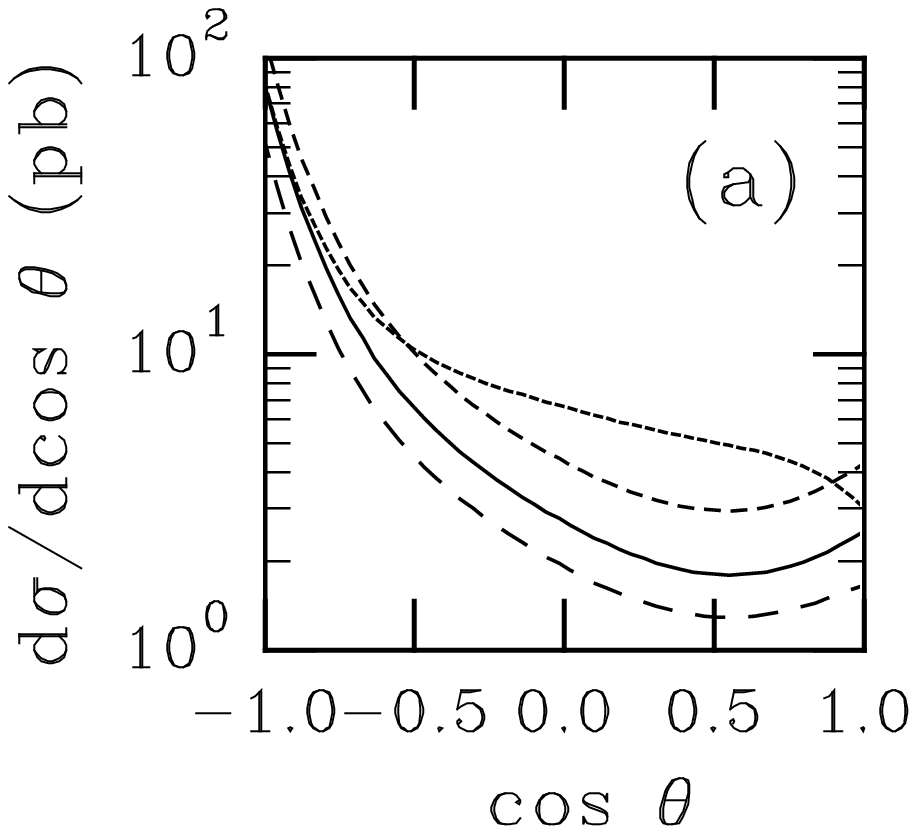,height=4cm}}
\end{center}
\end{minipage}
\hfill
\begin{minipage}{.495\linewidth}
\begin{center}
\mbox{\epsfig{figure=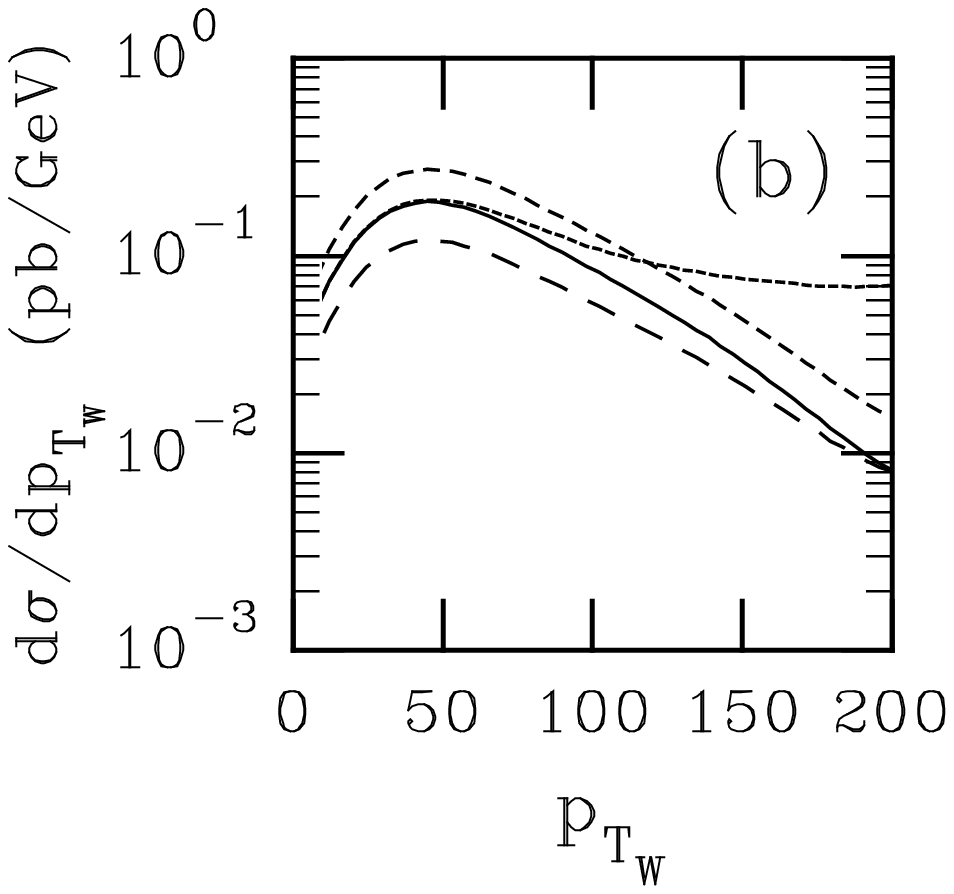,height=4cm}}
\end{center}
\end{minipage}
\noindent
{\small Fig. 2.  (a) The angular distribution of the
reconstructed $W$ boson
relative to the incoming electron.
(b) The $p_T$ distribution of the  reconstructed
$W$ boson.  The solid line is the standard model
prediction, the long-dashed line is for $\kappa_\gamma=0.6$,
$\lambda_\gamma=0$,  the short-dashed line is for $\kappa_\gamma=1.4$,
$\lambda_\gamma=0$,  and the dotted line is for $\kappa_\gamma=1$,
$\lambda_\gamma=0.4$.}
\end{figure}

\begin{figure}
\begin{center}
\mbox{\epsfig{figure=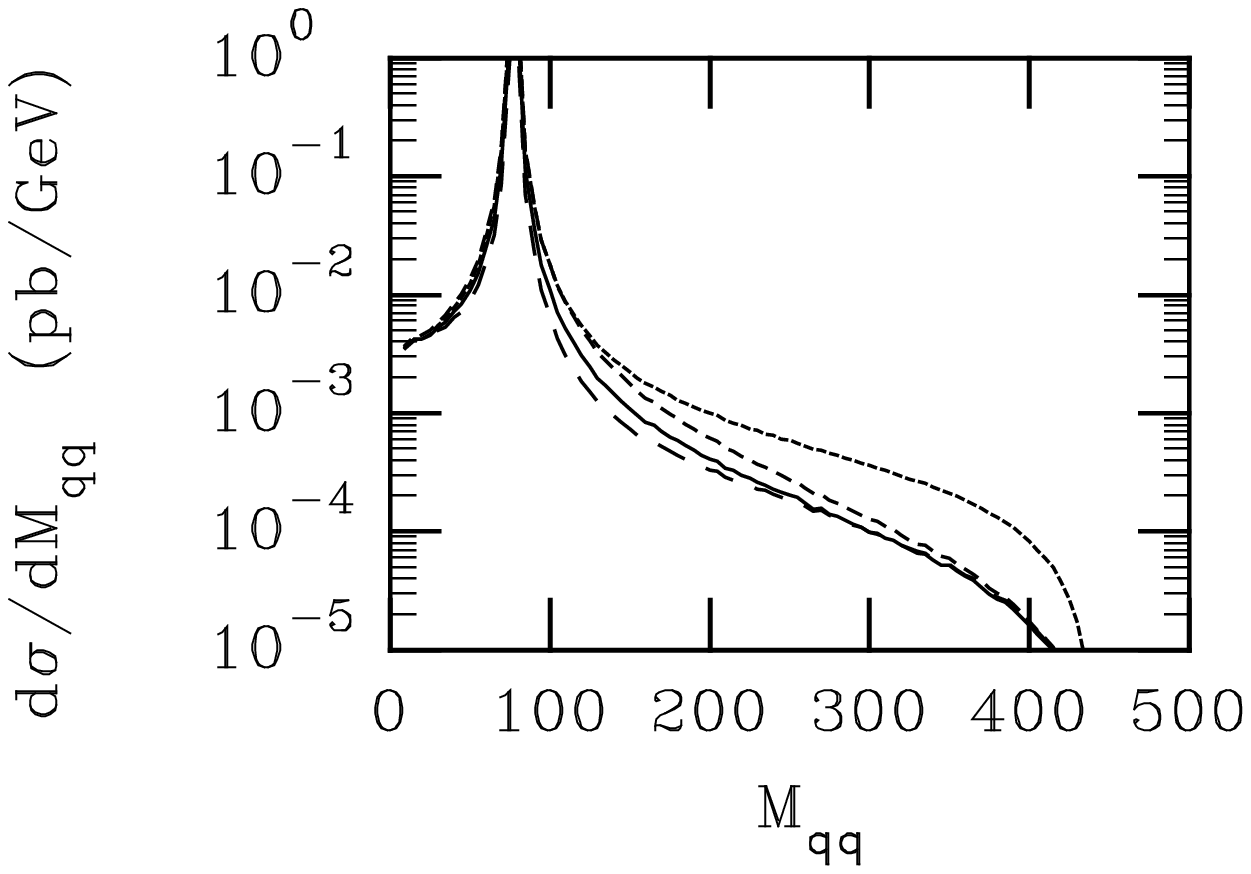,height=4.cm}}
\end{center}
\noindent
{\small Fig. 3. The hadron jet invariant mass ($M_{q\bar{q}}$)
distribution for $\sqrt{s}=500$~GeV.
The solid line is the standard model
prediction, the long-dashed line is for $\kappa_\gamma=0.6$,
$\lambda_\gamma=0$,  the short-dashed line is for $\kappa_\gamma=1.4$,
$\lambda_\gamma=0$,  and the dotted line is for $\kappa_\gamma=1$,
$\lambda_\gamma=0.4$. }
\end{figure}

To quantify the sensitivities of the TGV's to these observables we
binned the angular distributions into four
equal bins, divided the $p_T$ distributions into the 4 $p_T$ bins;
$0-100$~GeV, 100-150~GeV, 150-200~GeV, and 200-250~GeV,
and considered the hadronic cross section with $M_{q\bar{q}}> 300$ GeV.
The bounds obtained for these observables are shown in Fig. 4;
$\kappa_\gamma$ can be measured to within 7\% and
$\lambda_\gamma$ to within $\pm 0.05$ at 95\% C.L..

\begin{figure}
\begin{center}
\mbox{\epsfig{file=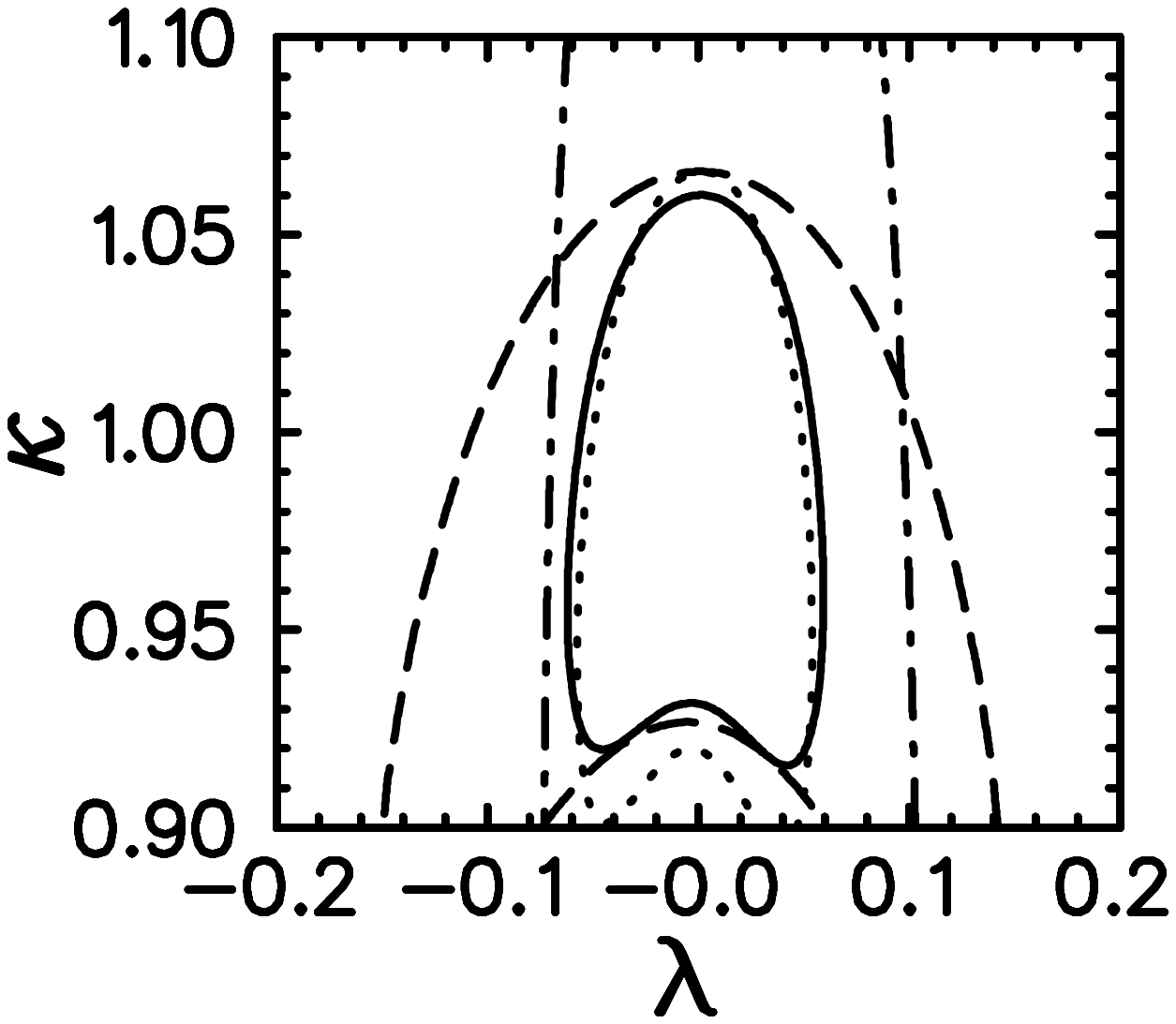,height=4.cm}}
\end{center}
\noindent
{\small Fig. 4.  The achievable bounds on $\kappa_\gamma$ and $\lambda_\gamma$
at 95\% C.L..
The dashed line is based on $d\sigma/d\cos \theta_W$,
the dotted line on $d\sigma/d {p_T}_W$,
the dot-dashed line is for $\sigma_{q\bar{q}}>300$
GeV, and the solid line is the combined angular and $p_T$ bounds.}
\end{figure}

\section{Conclusions}

We examined single $W$ production in $e\gamma$ collisions
for the  NLC 500 GeV $e^+e^-$ collider
using a backscattered laser photon spectrum.
We included the $W$ boson decays to final state fermions and other
processes which contribute to the same final state.
At high energy,  the off resonance results are important since
interference effects between these other diagrams and the $W$ production
diagrams enhance the significance of  anomalous couplings, particularly
$\lambda_\gamma$.  Although these effects contribute relatively little
to the total cross section, their significance in constraining the
anomalous couplings can be large, especially at high energies and high
luminosities where these effects are statistically significant.
We found sensitivities of
$\delta\kappa_\gamma \simeq \pm 0.07$ and $\delta\lambda_\gamma\simeq
\pm 0.05$.
The measurement of $\kappa_\gamma$
is approaching the level of radiative
corrections and might be sensitive to new physics at the loop level.
On the other hand, it is expected that the sensitivity to $\lambda$
would have to be at least an order of magnitude more sensitive to be
interesting.

\bibliographystyle{unsrt}

\end{document}